\documentstyle[twocolumn,aps,prb,epsf]{revtex}
\begin{document}
\title{
Stripes, Pseudogaps, and Van Hove Nesting 
in the Three-band tJ Model}

\author{R.S. Markiewicz} 

\address{Physics Department and Barnett Institute, 
Northeastern U.,
Boston MA 02115}
\maketitle

\begin{abstract}
Slave boson calculations have been carried out in the three-band tJ model for
the high-T$_c$ cuprates, with the inclusion of coupling to oxygen breathing mode
phonons.  Phonon-induced Van Hove nesting leads to a phase separation between a 
hole-doped domain and a (magnetic) domain near half filling, with long-range 
Coulomb forces limiting the separation to a nanoscopic scale.
Strong correlation effects pin the Fermi level {\it close to}, but
not precisely at the Van Hove singularity (VHS), which can enhance the tendency
to phase separation.

The resulting dispersions have been calculated, both in the uniform phases and
in the phase separated regime.  In the latter case, distinctly different 
dispersions are found for large, random domains and for regular (static) striped
arrays, and a hypothetical form is presented for {\it dynamic} striped arrays.
The doping dependence of the latter is found to provide an excellent description
of photoemission and thermodynamic experiments on  pseudogap formation in 
underdoped cuprates.  In particular, the multiplicity of observed gaps is
explained as a combination of flux phase plus charge density wave (CDW) gaps 
along with a
superconducting gap.  The largest gap is associated with VHS nesting.  The
apparent smooth evolution of this gap with doping masks a crossover from 
CDW-like effects near optimal doping to magnetic effects (flux phase) near half 
filling.  A crossover from large Fermi surface to hole pockets with increased
underdoping is found.
In the weakly overdoped regime, the CDW undergoes a quantum phase transition
($T_{CDW}\rightarrow 0$), which could be obscured by phase separation.
\end{abstract}


\narrowtext

\section{Introduction}

The slave boson technique has frequently been applied to the study of strong
correlation effects in metals\cite{slab,KLR}.  In the high-T$_c$ cuprate 
superconductors, the intense theoretical activity now allows a detailed 
comparison of slave boson results with the results of quantum Monte Carlo (QMC)
or exact diagonalization calculations.  Qualitatively, the 
comparison is excellent: the latter calculations have confirmed slave boson 
predictions\cite{KLR,sad,RM3,Gri,New} that (1) correlations preserve
a Fermi-liquid-like energy dispersion, but renormalize the bandwidth (`flat 
bands')\cite{flatb}; (2) a Mott transition to an insulating phase can only
occur at exactly half filling, when the bare Cu-O energy splitting $\Delta_0$ is
larger than a critical value $\Delta_{0c}$  
(compare Refs. \cite{Cast,RMXA} and \cite{EnHa}).  The 
agreement is semi-quantitative: the doping dependence of the chemical potential 
in the three-band model is nearly the same when calculated by slave boson or by 
quantum Monte Carlo techniques\cite{Dopf,Scal}, with a slightly smaller value of
$\Delta_{0c}$ in the latter (Fig. 36 of Ref. \cite{Surv}).
\par
The usual slave boson technique does not incorporate the magnetic effects which
are important near half filling.  This may be phenomenologically remedied by
including the lowest-order correction in $t^2/U\sim J$, producing a three-band 
tJ model\cite{DiDo,CaGr,CDG,Phas}.  A finite $J$ leads to a small reduction in 
$\Delta_{0c}$, thereby improving agreement with QMC calculations.  The present 
paper utilizes the three-band tJ model to explore the role of Van Hove 
singularity (VHS) nesting in the cuprates.  It is found that VHS nesting 
provides a natural explanation for the occurence of striped phases, composed of 
a magnetic-dominated regime near half-filling and a charge-dominated regime near
optimal doping.  The model is found to give a good description of the observed 
pseudogap formation in these materials, as well as to explain why the charged
stripes are pinned near the VHS.

\section{Self-consistent Equations}

In a recent survey of the Van Hove scenario\cite{Surv}, it was pointed out that 
there are actually two variants of the scenario, a {\it simple} and a {\it 
generalized} scenario.  The simple scenario explores the role of a peak in the 
density of states (dos) on the normal state and superconducting properties, 
ignoring any possible role of competing instabilities.  In the extended 
scenario, these competing instabilities -- predominantly spin or charge density 
waves -- play an essential role, which can lead to the suppression of the 
superconducting instability.  
\par
Prior slave boson calculations had shown that correlation effects (a large 
on-site Coulomb repulsion, $U$) tend to pin the Fermi level near the VHS over an
extended doping range\cite{RM3,New,QSi}, and this has since been confirmed by a 
number of different techniques\cite{inD,Pines,Dich,pinn,rgpin,MaKr}. It was 
suggested
that nesting of the VHS's would lead to density wave instabilities which could 
successfully compete with superconductivity.  This was modelled\cite{RM8a} in a 
one-band tight-binding model, where the pinning was approximated by adjusting 
the Fermi surface curvature such that the Fermi level was exactly at the VHS for
all dopings.  The resulting density wave-superconducting phase diagram is in 
good agreement with the experimentally observed pseudogap transition vs. 
doping\cite{RMPRL}.
\par
Clearly, such a model is oversimplified.  Strong correlation effects tend to
suppress electron-phonon coupling near half filling, suggesting that the density
wave must cross over from spin-density-wavelike near half filling to
charge-density wavelike near optimal doping.  Whether such a crossover could be
adequately described in a one-band model was not at all clear.
\par
In the present paper, a self-consistent three-band tJ model calculation is
presented for this crossover, confirming the main results of the simpler
calculation, and suggesting a possible origin of striped phases in these
materials.

The three-band tJ model Hamiltonian is the same as that presented in Ref. 
\cite{Phas}, with the addition of a term due to the formation of a
charge density wave (CDW) of breathing mode symmetry.  No spin-dependent terms
are included, so the present model does not describe either the N\'eel or the
superconducting phases.  Both the CDW and the
magnetic flux phase, if present, break the symmetry of the even and odd Cu
sites in a given layer, giving rise to a doubling of the unit cell area.  In
a basis set consisting of symmetric and antisymmetric combinations of the atoms
on the two sublattices, the Hamiltonian matrix becomes
\begin{equation}
\left(\matrix{\Delta_+&-2its_x&-2its_y&-i\Delta_m&
         -2tc_x\delta_b&-2tc_y\delta_b\cr
      2its_x&0&u_0s_xs_y&-2its_x\delta_a&0&0\cr
      2its_y&u_0s_xs_y&0&-2its_y\delta_a&0&0\cr
      i\Delta_m&2its_x\delta_a&
        2its_y\delta_a&\Delta_-&2tc_x&2tc_y\cr
      -2tc_x\delta_b&0&0&2tc_x&0&u_0c_xc_y\cr
      -2tc_y\delta_b&0&0&2tc_y&u_0c_xc_y&0
\cr}\right)
\label{eq:1}
\end{equation}
where $\Delta_{\pm}=\Delta\pm\Delta_p$.
In this matrix, the band parameters are $\Delta$, the splitting between the Cu 
and O energy levels, $t_{CuO}=t(1\pm\delta )$, the Cu-O hopping parameter, 
$t_{OO}$, the O-O hopping parameter, and $\Delta_1$ a parameter associated with 
Cu-Cu exchange.  In addition, $c_i=cos(k_ia/2)$, $s_i =sin(k_ia/2)$, $i=x,y$, 
and $u_0=-4t_{OO}$.
\par
$\delta$ is the asymmetry of the Cu-O hopping introduced by the
CDW distortion; for a breathing mode, all four Cu-O bonds of one Cu are long
($t_{CuO}=t(1-\delta )$), while all four bonds for the other Cu are short. 
There are two possible values for $\delta$: let
\begin{equation}
\delta_k=\cases{\delta_0+\delta_1,&if $|E_k^{\prime}-E_F|\le\hbar\omega_0$;\cr
                \delta_0,&otherwise,\cr}
\label{eq:2}
\end{equation}
with $\omega_0$ a phonon cutoff frequency and $E_k^{\prime}$ the quasiparticle 
energy in the absence of a CDW.  The use of $E_k^{\prime}$ in Eq. \ref{eq:2}
is a weak coupling approximation; it will be seen to lead to a slightly
erroneous dispersion, in that the resulting gap is not exactly centered at
$E_F$.  In the Hamiltonian matrix, Eq. \ref{eq:1}, the $\delta_i$ $i=a,b$ 
should be replaced by the appropriate $\delta_k$, which may be different for the
symmetric and antisymmetric Cu's.
\par
At the mean-field level, magnetic exchange leads to an effective Cu-Cu hopping,
of magnitude
\begin{equation}
\Delta_{ij}\equiv J\sum_{\sigma}<d_{i\sigma}d_{j\sigma}^{\dagger}>
=\Delta_1e^{i\theta_{ij}},
\label{eq:13d}
\end{equation}
with $i$ and $j$ labelling adjacent Cu sites.  A number of different magnetic
phases are possible, depending on the choice of phase.  Here only two 
magnetic phases are considered, the paramagnetic ($\theta_{ij}=0$) and the flux
($\theta_{ij}=\pm\pi /4$)\cite{Affl} phases.  The paramagnetic phase is usually
called the uniform phase, but here `uniform' will be used in a different way, to
denote the absence of a phase separation.  In the flux phase, the $\pm$ sign 
is chosen so that the net phase change around any plaquette is $\pm\pi$.  
For the paramagnetic phase, $\Delta_p=-2\Delta_1(\bar c_x+\bar c_y)$, 
$\Delta_m=0$, with $\bar c_i=cos k_ia$, $i=x,y$.  For the flux phase, 
$\Delta_p=-\sqrt{2}\Delta_1(\bar c_x+\bar c_y)$ and $\Delta_m=-\sqrt{2}\Delta_1
(\bar c_x-\bar c_y)$.
\par
If the unrenormalized values of $\Delta$ and $t$ are $\Delta_0$ and $t_0$
respectively, then setting $r_0=t/t_0$, the equations of self-consistency become
\begin{equation}
r_0^2={1\over 2}[1-{1\over N_s}\sum_ku_k^2f_h(E_k)],
\label{eq:3}
\end{equation}
\begin{equation}
\Delta_0-\Delta ={1\over 2r_0^2N_s}\sum_ku_k^2f_h(E_k)(E_k-\tilde\Delta ),
\label{eq:4}
\end{equation}
\vskip 0.9truein
and
\begin{equation}
\Delta_1={J\over 2N_s}\sum_ku_k^2f_h(E_k)\gamma_{\vec k},
\label{eq:5}
\end{equation}
where $N_s$ is the number of unit cells 
$E_k$ is the eigenvalue of $H$, $f_h(
E_k)$ is the Fermi function, $u_k$ is the amplitude of the wave function on Cu, 
and $\tilde\Delta =\Delta +2\Delta_1\gamma_{\vec k}$.  The function $\gamma_
{\vec k}=\bar c_x+\bar c_y$ ($\sqrt{\bar c_x^2+\bar c_y^2}$) in the paramagnetic
(flux) phase.  As written, Eqs.~\ref{eq:3} and \ref{eq:4} are valid for a hole 
picture, so $f_h(E_k)=1$ for $E_k>E_F$, and $=0$ otherwise (assuming $T=0$).  
The Fermi energy is determined from
\begin{equation}
{1\over N_s}\sum_kf_h(E_k)=1+x.
\label{eq:6}
\end{equation}

For the CDW, the additional self consistent equations are\cite{BFal}
\begin{equation}
\delta_0={-V_{ep}\over 2tN_s}\sum_k^{\prime\prime}
f_h(E_k)u_{1k}^*(v_{2x}c_x+v_{2y}c_y),
\label{eq:7}
\end{equation}
\begin{equation}
\delta_1={-V_{ep}\over 2tN_s}\sum_kf_h(E_k)u_{1k}^*(v_{2x}c_x+v_{2y}c_y),
\label{eq:8}
\end{equation}
where $V_{ep}$ is the phonon-induced effective electron-electron interaction 
energy and the double prime on the first sum means that both $|E_k^{\prime}-
E_F|\le\hbar\omega_0$ and $|E_{|\vec k+\vec Q|}^{\prime}-E_F|\le\hbar\omega_0$, 
with $\vec Q=(\pi /a,\pi /a)$.  Also, $u_1$ is the wave function amplitude for 
one (e.g., the symmetric) Cu, and $v_{2x}$, $v_{2y}$ are the corresponding 
amplitudes for the (antisymmetric) oxygens.  
\par
In the present paper, the parameters are taken as $\Delta_0=6eV$, $t_0=1.3eV$, 
$t_{OO}=-0.45eV$, $J=0.13eV$ and $\hbar\omega_0=50meV$\cite{Surv}.  The free 
energy is 
\begin{equation}
F=F_0+\sum_kf_e(E_k)E_k,
\label{eq:9}
\end{equation}
\begin{equation}
F_0=N_s\bigl[(\Delta_0-\Delta )(1+2r_0^2)+{2\Delta_1^2\over J}+{t^2\delta_0
\delta_1\over V_{ep}}\bigr].
\label{eq:10}
\end{equation}

\section{Phase Separation}

\subsection{Search for Phase Separation}

\par
There is considerable experimental evidence for phase separation in the 
cuprates, which has been presented in a number of conferences\cite{phassep} and 
reviews\cite{Eme1,RMrev,Surv}.  For the hole-doped cuprates, the phase 
separation is believed to be between a hole-doped phase and an antiferromagnetic
insulator (AFI) phase close to half filling.  The experimental evidence for this
latter case falls into two categories, 
depending on whether the dopant ions are mobile or not.  Thus, in La$_2$CuO$_
{4+\delta}$, the doping is provided by interstitial oxygens\cite{LCO} which are 
mobile below room temperature.  When the holes bunch up, the interstitial O's 
follow, leading to a macroscopic phase separation between an undoped AFI and an 
optimally doped high-T$_c$ superconductor.  In other cuprates, the dopant ions
are immobile, and the phase separation is restricted to a nanoscopic scale 
due to Coulomb repulsion between holes. Tranquada and coworkers\cite{Tran,Tran2}
have demonstrated that in La$_{2-x-y}$Nd$_y$Sr$_x$CuO$_4$, when $x\sim 1/8$,
commensurability effects pin the domains, allowing a clear observation of
alternating charged and magnetic stripes.  They suggest that similar stripes 
exist in other cuprates, but as dynamic fluctuations.
\par
There have been a number of theoretical suggestions that phase separation arises
in a doped Mott insulator, some\cite{AFI,Nag} prior to the discovery of 
high-T$_c$ superconductivity, and others\cite{HiSi,Eme,RM3} in the specific 
context of the cuprates.  However, detailed calculations have generally found
that phase separation is either absent (in the pure Hubbard model\cite{Su,ElDa})
or is present only for unphysically large choices of parameters, such as $J$ (in
the tJ model)\cite{ElDa}, or the nearest neighbor Coulomb energy $V$ (in the 
three-band extended Hubbard model)\cite{Rai}.  Some recent calculations of the
$tJ$ model have suggested that phase separation persists to lower values of 
$J$\cite{Erc,HM}, but these results remain controversial.  
\par
\begin{figure}
\leavevmode
   \epsfxsize=0.33\textwidth\epsfbox{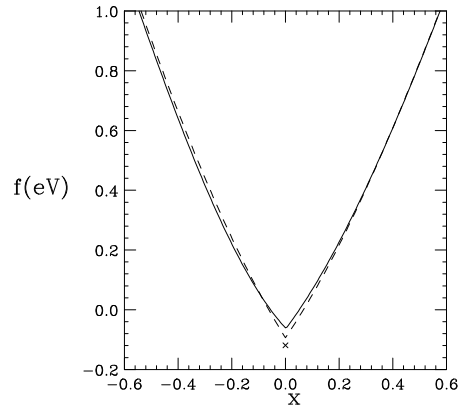}
\vskip0.5cm 
\caption{Comparison of free energies in the paramagnetic (solid line) and flux
(dashed line) phases.  To accentuate the curvature in $f$, a term linear in $x$ 
has been subtracted from all curves.  The $tJ$ model result at half filling is 
indicated by a $\times$.}
\label{fig:0a}
\end{figure}

The present calculations show no evidence of phase separation in the absence of
electron-phonon coupling.
Figure \ref{fig:0a} shows the doping dependence of the free energy for both the
paramagnetic (solid line) and flux (dashed line) phases, when 
$V_{ep}=0$\cite{Phas}.  The cusp at half filling is indicative of
the transition to a charge-transfer insulating state at half filling 
(discontinuity of the chemical potential).  At half filling, the flux phase is
more stable than the paramagnetic phase.  There is a crossover to the 
paramagnetic phase near $x=-.09$ for electron doping, or $x=0.38$ for hole
doping.  In the present calculation, 
these appear to be first-order transitions, but if the flux per plaquette is 
not restricted to the values zero and $\pi$ (the phase $\theta_{ij}$ in Eq. 
\ref{eq:13d} is allowed to vary continuously\cite{CaGr,CDG}), then there is
a smooth crossover from the flux to the paramagnetic phase, without a 
discontinuous jump (see Ref. \cite{gon}). 
\par
The present slave boson calculation underestimates the free energy at 
half filling, due to neglect of N\'eel order.  Since $t 
\rightarrow 0$ at half filling, the oxygen bands decouple from the problem, and 
the free energy should be identical to that found in the one-band $tJ$ model, 
and hence in the Hubbard model.  This energy is known to be $E_H=-0.66934J
$\cite{HM,Man}.  As denoted by the $\times$ in Fig. \ref{fig:0a}, this is $\sim
50\%$ lower than the flux phase result.  [This estimate neglects a term $-n_{Cu}
J/2$, which is common to all the magnetic phases and changes only weakly with
doping.  Here $n_{Cu}$ is the average hole density on the Cu.]
Since the N\'eel transition decreases rapidly with doping,
the free energy curve should cross over from the $\times$ to the dashed line at 
a fairly low doping value.  While there are other factors which can further 
lower the free energy near half filling, such as a spin-Peierls 
phase\cite{ZP,TH}, it does not appear that such effects will introduce a
tendency toward phase separation.  Since a cusp is already present at half
filling, it will be much more effective {\it to introduce a second free energy
dip away from half filling}.

\subsection{VHS Nesting Induced Phase Separation}

\begin{figure}
\leavevmode
   \epsfxsize=0.33\textwidth\epsfbox{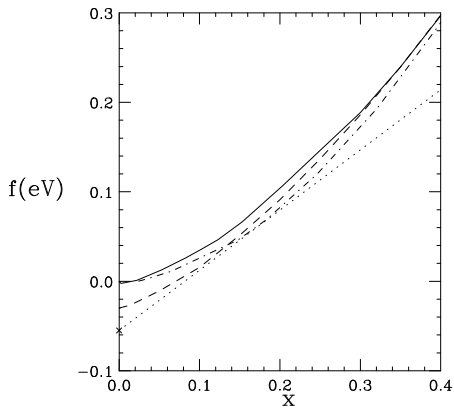}
\vskip0.5cm 
\caption{Free energy $f=F/N_s$ in the paramagnetic phase as a function of hole 
doping $x$, for electron phonon coupling $V_{ep}$ = 0 (solid line) or 1eV 
(dotdashed line), compared to the flux phase (dashed line). A term linear in $x$
has been subtracted from all curves.  The $tJ$ model result at half filling is 
indicated by a $\times$; dotted line = tangent construction for two-phase 
coexistence.}
\label{fig:1}
\end{figure}
\par
Figure~\ref{fig:1} compares the free energy $f=F/N_s$ for the 
paramagnetic phase for two values of $V_{ep}$ = 0 or 1eV to that of the flux
phase.  In the paramagnetic phase, a charge density wave (CDW) lowers the free 
energy by opening a gap at the VHS, lowering the electronic energy of occupied 
states. (The nature of the resulting CDW state is discussed further in the 
following section.)  Since the VHS degeneracy is already split in the flux 
phase (Fig. \ref{fig:6} below), no additional energy lowering is possible, and 
the CDW is not compatible with the flux phase. From Fig.~\ref{fig:1}, it can be
seen that the free energy lowering due to the CDW has a strong $x$-dependence: 
the free energy lowering vanishes as $x\rightarrow 0$ and is absent for electron
doping.  It also vanishes for too large hole doping.  This result is due to VHS
nesting: the strongest CDW effects occur when the Fermi level is close to the 
VHS.  
\par
These results confirm and quantify the prediction\cite{RM3} that electron-phonon
coupling near the VHS produces a dip in the free energy, which can generate 
a phase separation.  The curves of free energy for the CDW phase and the flux
phase cross at a finite hole doping, $x$.  Since it is not possible for the
system to smoothly evolve between the two phases, this indicates a first order
phase transition, with two-phase coexistence regime given by a tangent 
construction.  In Fig.~\ref{fig:1}, it is assumed that the phase separation
starts from the AFI phase at half filling (denoted by $\times$), but
depending on the exact dispersion, the free energy minimum may be shifted 
off of half filling.  The metallic phase will tend to be pinned near the VHS, as
discussed further in the following section.  
\par
Note that the present calculation is perhaps the strongest indication to date
that phase separation can arise in the cuprates with a realistic choice of
band parameters.
\par
In the following, it will be assumed that there is a phase separation between 
a flux phase with $x=
x_1$ and a CDW phase with $x_c\simeq 0.2$.  The precise value of $x_c$ 
is not important, but it will turn out to make a difference whether the flux 
phase is at $x_1=0$ or $0^+$.  Before this phase separation is analyzed, the
properties of the uniform phases will be briefly discussed.

\section{Results: Uniform Phases}

\subsection{CDW Gap}

Whereas equations Eqs.~\ref{eq:7} and \ref{eq:8} represent a BCS-like 
calculation of the CDW gap\cite{BFal}, the nature of the gap is very different 
from that found in a superconductor, since the pairing now involves an electron 
and a hole.  From Equation~\ref{eq:2}, there are {\it two gaps} with very
different properties.  The term $\delta_0$ produces a uniform gap throughout the
Brillouin zone.  This gap is {\it not} tied to the Fermi level, but has its own 
dispersion throughout the zone.  However, it {\it is} tied to the VHS, and
always splits the VHS density of states (dos) peak into two components.  On the
other hand, the gap associated with $\delta_1$ is localized near the Fermi
level, but need not split the VHS degeneracy.  
\par
In the cuprates, it will be shown that the gaps near the VHS's tend to
change the large Fermi surface into pockets near the $(\pi /2,\pi /2)$ points.
Since the $\delta_1$ gap only acts to enhance the $\delta_0$ gap, the Fermi 
surface near these pockets remains ungapped, and hence available for,
e.g., superconducting pairing.  
\begin{figure}
\leavevmode
   \epsfxsize=0.33\textwidth\epsfbox{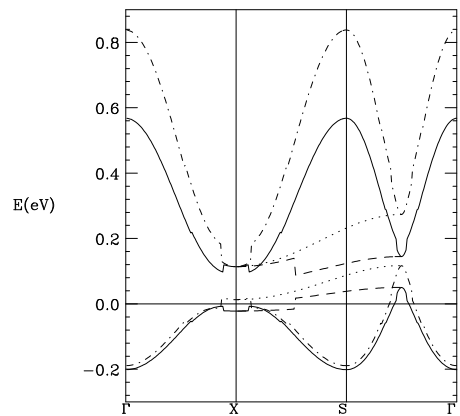}
\vskip0.5cm 
\caption{Energy dispersion for the paramagnetic phase, for $V_{ep}$ = 1eV, for 
$x$ = 0.15 (solid lines) or 0.32 (dotdashed lines).  The dashed and dotted lines
are the corresponding dispersions from $X$ to $\bar S=S/2$.}
\label{fig:0}
\end{figure}
\par
These features are illustrated in Figure~\ref{fig:0}, for two different dopings 
with $V_{ep}$ = 1eV.  For both dopings, $\delta_0$ leads to a similar splitting 
of the VHS degeneracy, but the $\delta_1$-associated gap is very different.  For
$x$ = 0.32, the Fermi level is more than $\hbar\omega_0$ below the VHS, so there
is a larger gap away from the VHS.  For $x$ = 0.15, the VHS is now within
$\hbar\omega_0$ of the Fermi level, but now for both of the bands coupled by
nesting (i.e., in Eq. \ref{eq:1}, $\delta_a=\delta_b=\delta_0+\delta_1$), so 
there are actually two $\delta_1$-type gaps near the VHS.  Note that the gap 
associated with $\delta_1$
is not exactly centered at $E_F$, and indeed has some dispersion of its own.  
This is due to the weak coupling Eq. \ref{eq:2}, which measures the gap from
the bands {\it in the absence of electron-phonon coupling}.
\par
When Eq.~\ref{eq:1} is Fourier transformed back to real space (and deconvolved 
from a symmetric/antisymmetric basis to an atomic basis), the term $\delta_a+
\delta_b$ is found to correspond to a uniform breathing mode distortion, while 
$\delta_a-\delta_b$ produces an additional modulation with periodicity $\sim 
k_F^{-1}$, where $\vec k_F$ is the wave number at the enhanced gap.  This can be
understood on the basis of simple hole counting: The uniform breathing 
mode distortion causes a doubling of the unit cell area.  This can only 
produce a gap at exactly half filling (corresponding to a filled band in the
supercell).  To produce a gap at $x\ne 0$ requires a large superlattice, as
would be produced by a commensurate value of $\vec k_F$.
\par
These results clarify an issue that had been raised earlier\cite{RMXB}. Whereas 
CDW effects have traditionally been associated with Fermi surface nesting, it 
was pointed out that polaronic band narrowing effects can be pinned to the VHS, 
and not the Fermi level, and hence can explain the observation of {\it
extended} VHS's\cite{exten}.  Here we see that there are three related effects.
The $\delta_0$ gap is associated with the polaronic effects tied to the VHS,
and will be seen to describe the photoemission observations of extended VHS's.
In addition, there is an extra gap associated with $\delta_1$, when the 
quasiparticle energy is close to the Fermi level.  Finally, this latter gap has
a nesting enhancement when two pieces of Fermi surface are separated by the
nesting vector (here, $Q=(\pi /a,\pi /a)$).  From Eq.~\ref{eq:8}, only in the
latter case does the gap contribute to the self-consistent equation for 
$\delta_1$.

\subsection{VHS Pinning}

\begin{figure}
\leavevmode
   \epsfxsize=0.33\textwidth\epsfbox{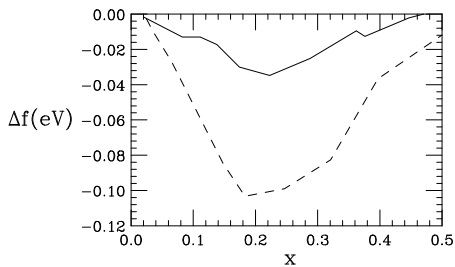}
\vskip0.5cm 
\caption{Free energy difference $\Delta f=f(V_{ep})-f(0)$ for the paramagnetic
phase with $V_{ep}$ = 0.6eV (dashed line) and 1.0eV (dotdashed line).}
\label{fig:2}
\end{figure}
\par
\begin{figure}
\leavevmode
   \epsfxsize=0.28\textwidth\epsfbox{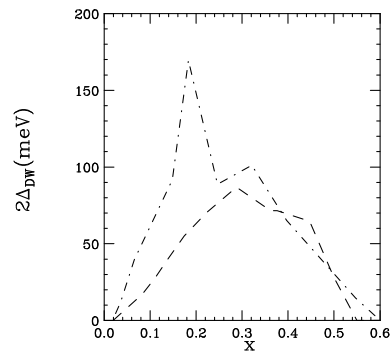}
\vskip0.5cm 
\caption{CDW gap $2\Delta_{DW}$ plotted vs doping in the paramagnetic phase for 
$V_{ep}$ = 0.6eV (dashed line), or 1eV (dotdashed line).}
\label{fig:4}
\end{figure}

Figure~\ref{fig:1} shows a crossover in free energy between the flux phase near
half filling and the CDW phase near the VHS, which leads to a regime of phase
separation.  In this subsection, the doping dependence of the CDW phase is
discussed in the absence of phase separation.  Since the CDW couples to the
electronic subsystem via modulation of the hopping parameter, the free energy 
lowering due to the CDW vanishes as $x\rightarrow 0$, where correlation effects
drive $t\rightarrow 0$.  When the hole doping gets too large, the Fermi level
moves beyond the VHS, and the stabilization energy also vanishes.  Hence, the 
strongest CDW effects occur when the Fermi level is close to the VHS (VHS 
nesting). Note that, since the {\it shape} of the Fermi surface is doping 
dependent, the VHS remains pinned close to the Fermi level over an extended 
doping range.  Nevertheless, the free energy lowering has a well defined 
maximum.  This is better seen in Fig.~\ref{fig:2}, which plots the difference in
free energy between the calculations for finite $V_{ep}$ and those with $V_{ep}
=0$.  This free energy lowering is approximately quadratic in $V_{ep}$, with a
doping dependence which roughly follows the magnitude of the CDW gap, 
$\Delta_{DW}$, Fig. \ref{fig:4}.
\par
\begin{figure}
\leavevmode
   \epsfxsize=0.33\textwidth\epsfbox{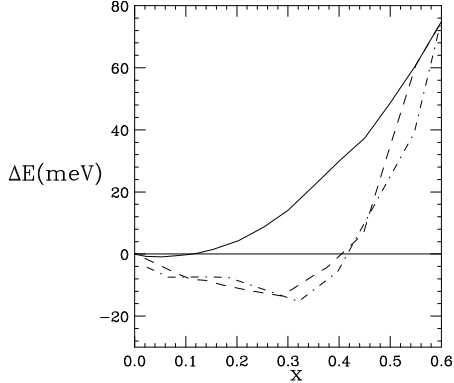}
\vskip0.5cm 
\caption{Energy separation between the VHS and the Fermi level $\Delta E=E_{VHS}
-E_F$, in the paramagnetic phase for $V_{ep}$ = 0 (solid line), 0.6eV (dashed 
lines), or 1eV (dotdashed lines).  For $V_{ep}\ne 0$, the VHS is split; the
present lines refer to the lower VHS's.  For the $V_{ep}=1$eV data, the extra
splitting due to $\delta_1$ is neglected.}
\label{fig:3}
\end{figure}
\begin{figure}
\leavevmode
   \epsfxsize=0.33\textwidth\epsfbox{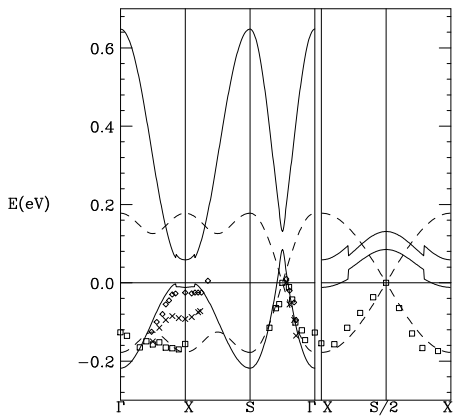}
\vskip0.5cm 
\noindent
\caption{Energy dispersion for the flux phase at half filling $x_1=0^+$ (solid 
line) and the paramagnetic phase at $x=$ 0.22, V$_{ep}$ = 0.6eV (dashed line),
close to the minimum of $\Delta f$.  Data from underdoped Bi-2212 (diamonds and 
$\times$'s)\protect\cite{Gp0} or SCOC (squares)\protect\cite{Well} are plotted 
as E/2.  Special points of the Brillouin zone are X = ($\pi$,0), S = ($\pi ,\pi
$).}
\label{fig:6}
\end{figure}

The CDW formation also greatly enhances the pinning of the Fermi level to the 
VHS.  Figure~\ref{fig:3} plots 
the energy difference between the VHS and the Fermi level.  In the absence of
electron-phonon coupling (solid line), the Fermi level crosses the VHS twice, 
at $x=0$ and $x=\tilde x_c$, and correlation effects pin the Fermi level close 
to the VHS for $0\le x\le\tilde x_c$.  However, for the large assumed value $t_
{OO}=-.45eV$, $\tilde x_c$ is small, leading to pinning in a narrow range of 
doping only. In the presence of a CDW, the energy bands split near the $X$-point
of the Brillouin zone, Fig. \ref{fig:6}, with each $X$-point band edge forming a
separate VHS dos peak.  The Fermi level is now even further from the average 
position of the VHS, taken as the middle of the $X$-point gap.  However, the 
{\it lower VHS peak is found to be pinned about 10-20meV below the Fermi level 
over a wide range of dopings}, Fig. \ref{fig:3}.  This is in striking agreement 
with photoemission observations of optimally doped cuprates.  Due to this strong
pinning effect, it is difficult to define just what doping corresponds to {\it
the} VHS.  However, the hole-doped end phase of the phase separation regime
will almost certainly be found to be pinned near a VHS.
\par
These electron-phonon effects are found to be absent in the flux phase.  This is
because the flux phase itself has already taken advantage of VHS nesting to
lower its free energy, as can be seen in Fig. \ref{fig:6}.
While there is no evidence for long-range CDW order in the cuprates, there is
considerable evidence for short-range lattice disorder, summarized in Ref. 
\cite{Surv}, Section 9.2.  While a substantial part of this local order is 
associated with tilting the CuO$_6$ octahedra out of the planes, there is also a
significant contribution associated with CuO bond stretching\cite{RMXB}.  
Considerable work will be required to sort out the relative contributions of 
various phonon modes.  For now, the breathing mode CDW is chosen to 
approximately represent the free energy lowering associated with this 
short-range order.

\subsection{Insulating Regime}

At half filling, for $\Delta_0>\Delta_{0c}$, there is an insulating phase with
$r_0=0$, but $\Delta_1\ne 0$, with energy dispersion 
\begin{equation}
E=\Delta+2\Delta_1\gamma_{\vec k}.
\label{eq:11}
\end{equation}
At this doping, the free energy has a cusp (Fig. \ref{fig:1}), so the chemical
potential has a discontinuity: $\Delta$ in Eq. \ref{eq:11} takes on different
values depending on whether $x=0$ is approached from positive or from negative
values.  These two states have the same free energy, even though their effective
Fermi levels differ by the charge transfer gap.  Hence the chemical potential is
pinned in the middle of the gap, and photoemission should see a band with finite
dispersion, Eq. \ref{eq:11}, separated from the chemical potential by half the
charge transfer gap.  

Note that this band is only half full, but because of strong correlation effects
the remaining states are no longer accessible. This is reminiscent of the upper
and lower Hubbard bands, although these are now charge-transfer bands.  This 
splitting of the band is readily done in the flux phase, since the lower and
upper halfs of the band are only connected at a few isolated points (diabolical
points).  However, it creates topological problems for the paramagnetic phase. 
Since the Fermi surface in the paramagnetic phase is square (at exactly half
filling, Eq. \ref{eq:11}), these problems can easily be overcome by opening a 
gap at the Fermi level -- either due to antiferromagnetism or to a period 
doubling charge density wave.  Note that this strongly suggests that the opening
of a correlation gap must be accompanied by some other kind of ordering.

The split-off charge transfer band may have been observed 
experimentally\cite{Well}. Since $t\rightarrow 0$, the dispersion is independent
of all hopping parameters (e.g., $t_{OO}$).  Hence, in the absence of 
longer-range exchange terms, the dispersion should be characteristic solely of 
the type of magnetic order, and should be the same as in the one band tJ model.
Thus, it is interesting to note that the dispersion in the paramagnetic phase 
matches that calculated in the tJ model\cite{DagN,Naz1}, while the flux phase 
dispersion matches that found in insulating Sr$_2$CuO$_2$Cl$_2$ 
(SCOC)\cite{Well}, Fig.~\ref{fig:6}.  This is consistent with the results of
Laughlin\cite{Lau1} and Wen and Lee\cite{WeL}.
\par
The magnitude of the predicted bandwidth can also be estimated.
Using the mean field decoupling, the equilibrium value of $\Delta_1$ in the
paramagnetic phase is $4J/\pi^2=0.406 J$, which is comparable to that found
in Ref. \cite{DagN}, $\Delta_1\simeq 0.55J$.  In the flux phase, $\Delta_1=0.479
J$, so $f=-2\Delta_1^2/J=-0.459J$, considerably smaller than the one band $tJ$
result, $E_H=-0.66934J$.  As discussed above, this is presumably due to 
neglecting the N\'eel order, and will be addressed in a future publication.  
$\Delta_1$ monotonically decreases in magnitude as the system is doped away from
half filling (in agreement with four-slave-boson calculations\cite{TIF}).  

\subsection{Flux Phase and VHS Nesting}

Within the present model, the flux phase has what is interpreted as a gauge 
degree of freedom\cite{Affl,Wang0,Surv}: the energy does not depend on the
individual phases $\theta_{ij}$ in Eq.~\ref{eq:13d}, as long as the sum of the
phases around a plaquette is $\pi$.  However, different choices of $\theta_{ij}$
lead to {\it different electronic dispersions}.  Since the electronic dispersion
is observable (e.g., by photoemission), this cannot be simply a choice of gauge.
For instance, the present choice (all phases equal in magnitude to $\pi /4$) 
leads to an `extended s-wave' type gap, with zero gap at the four points 
equivalent to $(\pi /2,\pi /2)$ and a maximum gap at the $(\pi ,0)$-type points;
this is equivalent to a VHS nesting gap.  Alternatively, concentrating the full
phase $\pi$ on a single bond (with $\theta_{ij}=0$ on the other three bonds)
leads to gap zeroes at the $(\pi ,0)$ points, and maxima at the $(\pi /2,\pi 
/2)$ points -- corresponding to conventional (flat Fermi surface) nesting --
see Fig. 32b,c in Ref. \cite{Surv}.  Both phases have the same energy, and yet
experimentally only the former is observed, Fig.~\ref{fig:6}.
\par
It is the discreteness of the CuO$_2$ lattice which breaks the gauge symmetry --
e.g., in photoemission, the position of the $\Gamma$ point is well defined.
Similarly, it is presumably a structural distortion (spin-Peierls-like effect)
which breaks the energy degeneracy, and locks in a particular distortion
pattern.  This will be explored further in a future publication.

\section{Pseudogaps in the Underdoped Regime}

\subsection{The Experimental Situation}

\par
Above, it was shown that the CDW gap is expected to have two components, one
tied to the VHS and one to the Fermi surface.  In a similar fashion, 
photoemission studies in underdoped Bi$_2$Sr$_2$CaCu$_2$O$_8$ (Bi-2212) find 
two gap-like features, one
tied to the Fermi level, and the other much larger gap near the VHS.  The 
{\bf small pseudogap} resembles a superconducting gap, with the photoemission
intensity collapsing to zero for $|E-E_F|\le\Delta_{sm}(\vec k)$; the {\bf
large pseudogap} is a shift of a broadened quasiparticle-like peak away from
the Fermi level, predominantly near $(\pi ,0)$ and $(0,\pi )$.  The small
pseudogap has a d-wave-like symmetry, with a maximum value of 25meV near $(\pi 
,0)$ and a
minimum value of $\sim 0$ near $(\pi /2,\pi /2)$\cite{Gp1,Gp2}.  The gap 
magnitude is nearly independent of doping, but it opens up at the pseudogap 
temperature -- i.e., at the superconducting transition in optimally doped 
material, but at a higher temperature in underdoped samples.  This gap thus has 
some features of the Fermi surface CDW gap discussed above, but combined with 
the superconducting gap.  This near Fermi surface feature will be analyzed 
further in the discussion section.  Here, I would like to concentrate on the
large pseudogap.
\par
In the underdoped regime, the photoemission studies find a peak near the $X$
point, which shifts further below the Fermi level with increasing 
underdoping\cite{Gp0}, Fig. \ref{fig:6}.  The Stanford group\cite{Gp0} reports a
continuous evolution of this peak with underdoping, approaching a dispersion
near half filling similar to that found in the insulating compound 
SCOC\cite{Well}, squares in Fig.~\ref{fig:6}.  
\par
On the other hand, Campuzano\cite{Camp} proposes a different doping dependence 
of this feature. He suggests that there are two independent features near $(\pi 
,0)$ in the Brillouin zone, one a sharp quasiparticle peak which is near the 
Fermi level at optimal doping, and broadens severely in underdoped samples, and 
the second a broad peak which is already present at $\sim$200meV below the Fermi
level in optimally doped material, and gradually shifts to 300meV with increased
underdoping.  The present model suggests an intermediate position: the data can
be understood as a dynamic average of the two separated phases represented
by the solid and dashed lines in Fig. \ref{fig:6}.  If the fluctuations are fast
compared to the experimental observation technique, the data will give an
average dispersion which evolves smoothly with doping.  A quasistatic 
fluctuation, on the other hand, would produce two coexisting peaks, with one 
peak gradually disappearing as the other peak grows up with underdoping.
\par
Thermodynamic measurements of the pseudogap by Loram, et al.\cite{Gp3,Gp6} in 
underdoped YBa$_2$Cu$_3$O$_{6+y}$ (YBCO) and LSCO are 
in good agreement with the Stanford photoemission data\cite{Gp0}.  They have 
measured the dos from susceptibility and heat capacity measurements and find 
that a pseudogap appears and grows with successive underdoping.  At $y=0.3$ in
YBCO, the peak of the gap is at $\Delta_g\simeq 100meV$, comparable to the 
larger photoemission gap seen in Bi-2212.  The gap can be fit to a d-wave gap --
i.e., a logarithmic divergence at $E=E_F+\Delta_g$, as at a VHS, but with the 
dos$\rightarrow 0$ at $E\rightarrow E_F$.  Presumably, these measurements are 
seeing a superposition of the two gaps seen in photoemission.
\par
There are also photoemission data on underdoped YBCO\cite{Liu,Liu2}, but they do
not reveal a similar gap opening.  Since the effects are rather subtle, 
additional measurements may be needed.

\subsection{Pseudogaps in the Uniform Phases}

Figure \ref{fig:6} compares the dispersion observed for 
a series of underdoped samples of Bi$_2$Sr$_2$Ca$_{1-x}$Dy$_x$CuO$_8$ with the 
theoretical dispersions from the two equilibrium phases at $x=0^+$ and $x=x_c$.
For the present choice of band parameters, the bandwidth is 
underestimated by about a factor of two, but the theoretical calculations match 
the overall dispersion in the optimally doped and extremely underdoped limits.
Indeed, as far as the large pseudogap is concerned, its doping dependence is
fairly well explained by the fact that it evolves into the flux phase at half
filling.

\begin{figure}
\leavevmode
   \epsfxsize=0.33\textwidth\epsfbox{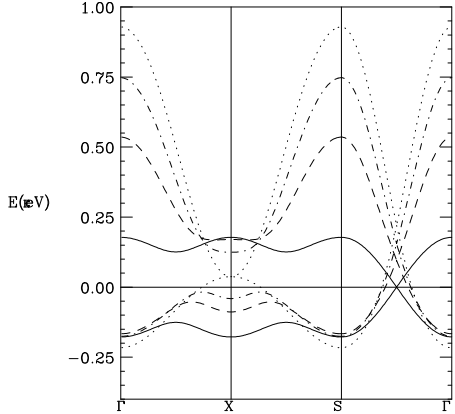}
\vskip0.5cm 
\noindent
\caption{Energy dispersion for the flux phase at $x$ = $0^+$ (solid line), 0.15 
(dashed line), and 0.30 (dotdashed line), and the paramagnetic phase at $x$ =
0.45 (dotted line).}
\label{fig:6z}
\end{figure}

Thus, figure~\ref{fig:6z} shows the energy dispersion in a series of uniform 
(flux or paramagnetic) phases, corresponding to the free energy curves in 
Fig.~\ref{fig:0a}.  As the flux phase is doped away from half filling
(perfect nesting), the $X$-point gap gradually closes.  The transition to the
paramagnetic phase occurs when the lower VHS gets too close to the Fermi level.
Note that the doping dependence already does a good job of reproducing the
photoemission data on the large pseudogap (consistent with the results of 
Preuss, et al.\cite{PHGE}), but cannot reproduce the small, $\sim 25meV$ 
pseudogap.  On the other hand, when $V_{ep}\ne 0$, the paramagnetic phase near 
optimal doping clearly displays a gap consistent with the small pseudogap, 
Fig.~\ref{fig:6}.  However, this agreement breaks down at intermediate 
dopings.  Figure~\ref{fig:4} shows that the density wave gap decreases with 
decreasing $x$, while the experimental gap increases.  Moreover, since the 
two pseudogaps have independent origins -- a CDW or a flux phase -- a theory
involving uniform phases cannot explain their coexistence. In the following 
section, it will be shown that
dynamic phase separation allows a smooth evolution between the two limits.

\section{Calculations for Striped Phases}
\par
In the presence of phase separation, the dispersion will change.  For a
macroscopic phase separation, the photoemission would be a superposition of
the two coexisting phases.  However, in the present case, the phase separation
is due to the holes only, so due to strong Coulomb effects (charging of the
domains), the ensuing phase separation is on a nanoscopic scale only, which
should lead to a more complex dispersion.  There is evidence that in the
cuprates, the phase separation is manifested in the form of alternating charge
and magnetic stripes\cite{Tran,Tran2}, with a well defined periodicity that 
varies smoothly with doping.  For any commensurate periodicity, the dispersion
will have additional minigaps associated with the superlattice 
periodicity\cite{RMIX}.  Since the stripes are generally dynamic, these 
superlattice gaps will probably be washed out.
\par
A proper calculation of the self-consistent, dynamic stripe phases is beyond the
scope of the current paper.  Instead, I will suggest a number of plausible
forms for the average dispersion, and show that it is possible to explain the
observed photoemission data.  There is considerable flexibility in the results,
and it appears likely that different dispersions can be observed, depending on 
the spacing of the stripes, and on whether the stripes are effectively static or
dynamic (i.e, on the time scale of the observational technique).

\subsection{Large Stripes}

To simplify the problem, it is assumed that the fluctuations lead to a disorder
in which any local region could have either of the two dispersions at random.
The average dispersion can then be calculated by a CPA (coherent phase 
approximation) calculation\cite{EKL,RMIX}.  Repeating the calculation of Ref.
\cite{RMIX}, but assuming a random mix of only two phases, with Green's 
functions $G_1$ and $G_2$ and probabilities $p_1$ and $p_2=1-p_1$, the average
Green's function is (exactly) given by 
\begin{equation}
G_0=p_1G_1+p_2G_2.
\label{eq:12}
\end{equation}
(Note that the assumption that the domains are large enough to have well
defined $G_1,G_2$ is an implicit assumption of macroscopic phase separation.)
Since the dispersion is given by real solutions of $G_0^{-1}=0$, the 
photoemission dispersion should simply be a weighted superposition of the
dispersions of the two end phases, with $G_1^{-1}=0$ or $G_2^{-1}=0$.  (The
latter correspond to the eigenvalues of the Hamiltonian, Eq.~\ref{eq:1}, in the
two coexisting phases.)  A similar solution was found in Refs. \cite{Kot,Surv}.
\par
Here, the distinction between the flux phase being at $x=0$ or $0^+$ can be
readily understood.  At $x=0$, the Fermi level lies in the middle of the charge
transfer gap.  Hence, the doping dependence predicted by Eq. \ref{eq:12} implies
a gapped state at half filling, which persists in the doped material, with the 
appearence of midgap states in the doped material, with intensity increasing 
linearly with $x$.  This is the form discussed in Refs. \cite{Kot,Surv}, and
seems to provide a good description of photoemission experiments in a number of
three-dimensional d-electron systems\cite{Mottran}.  It would also be
consistent with the photoemission data on underdoped YBCO\cite{Liu}, although
it is not clear why these data are inconsistent with the heat capacity
results\cite{Gp3}.
\par
However, this does not provide a good description of the photoemission data in
Bi-2212, Fig. \ref{fig:6}.  A simple modification of the theory can 
significantly improve the agreement.  If the phase separation starts not at $x=0
$, but at a small positive doping $x_1=0^+$, then the doping dependence of the
photoemission would be as follows.  For doping between $x=0$ and $x_1$, there
is no phase separation.  The system remains in the flux phase, with the Fermi
level again at midgap in undoped material, but shifting immediately to the top
of the charge transfer band as soon as the first holes are doped in.  This
state is in good agreement with the experimental data on the most underdoped 
Bi-2212 samples, Fig. \ref{fig:6}.  With increased doping, the two phase regime 
is reached.  (Note that in this doped regime, the optical conductivity can still
see a charge transfer gap\cite{KLR}.)
\par
The present model would predict a superposition of the two
end phases, rather similar to Campuzano's interpretation of the data, but not
consistent with the continuous evolution of the pseudogap suggested by the
Stanford\cite{Gp0} and Loram\cite{Gp3,Gp6} results.
This lack of agreement is not surprising.  Such a macroscopic phase separation
would also predict a unique value of the superconducting transition temperature
$T_c$ (the flux phase at $x_1$ being nonsuperconducting).  The experimental
observation that $T_c$ evolves smoothly with doping strongly suggests that the
phase separation is on such a nanoscopic scale that $T_c$ evolves with doping
via a form of proximity effect between the two phases.  This same effect should
explain the observed behavior of the pseudogap.
\par
On a deeper level, the striped phases constitute a new thermodynamic state of
matter, which can have a doping dependent $T_c$.  A good analogy is provided by
a superconductor in a magnetic field.  In a type I superconductor, there is a
macroscopic phase separation between domains wherein the magnetic field is non
zero, quenching the superconductivity, and superconducting domains with zero 
field.  Increasing the field reduces the fraction of the material which is 
superconducting, but $T_c$ does not change with field.  In a type II
superconductor, the field domains are shrunk down to nanoscopic size as
magnetic vortices, producing a novel state of matter in which $T_c$ is a
continuously varying function of the field.  In the following subsections, a
model is provided for the pseudogap in the striped phases.

\subsection{Small Static Stripes}

If the stripe pattern is static and commensurate with the crystalline lattice,
then the stripes will provide a superlattice to modify the dispersion of the
separate phases.  Since the resulting dispersion is profoundly modified, it
would be appropriate to repeat the self-consistent calculations in the presence
of the stripes.  Without attempting this difficult procedure, however, a
qualitative understanding of the results can be achieved by using the
self-consistent band parameters calculated for the two phases in equilibrium,
at $x=x_1>0$ and $x=x_c$.  For definiteness, the end phases are assumed to be
the flux phase at $x_1=0^+$ and the paramagnetic phase with $V_{ep}=0.6eV$, $x_c
=0.288$.
\begin{figure}
\leavevmode
   \epsfxsize=0.33\textwidth\epsfbox{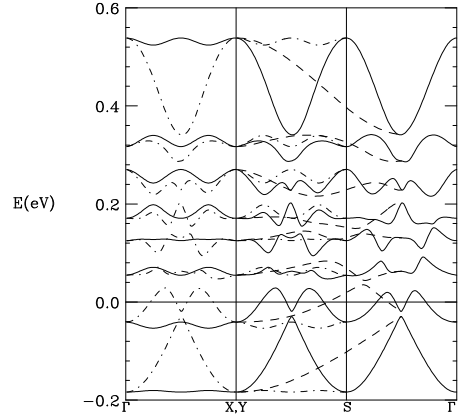}
\vskip0.5cm 
\caption{Energy dispersion for a static striped phase, with $n=2$ charge layers 
and $m=2$ magnetic layers.  Solid line (dotdashed line): dispersion along 
$\Gamma\rightarrow X(Y)\rightarrow S$; dashed line: dispersion along $X
\rightarrow S/2$.}
\label{fig:7a}
\end{figure}
\begin{figure}
\leavevmode
   \epsfxsize=0.33\textwidth\epsfbox{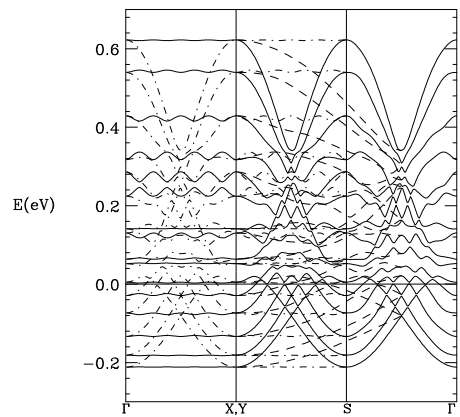}
\vskip0.5cm 
\caption{Energy dispersion for a static striped phase, as for Fig. 
\protect\ref{fig:7a}, but with $n=6$ charge layers 
and $m=2$ magnetic layers.}
\label{fig:7b}
\end{figure}
\par
Specifically, the stripes are assumed to be uniform along the $y$ direction, and
of periodicity $m+n$ Cu atoms along $x$, with $m$ Cu atoms in the magnetic phase
and $n$ Cu's in the charged phase ($x=x_c$).  For the cuprates, the $x$ and $y$
axes run parallel to the Cu-O-Cu bonds.  All of the band parameters can be 
assigned values corresponding to either the flux phase at $x_1$ or the 
paramagnetic phase at $x_c$, except for the magnetic coupling of a Cu atom in 
the magnetic phase with a neighboring Cu in the charged phase.  For these atoms,
$\Delta_1$ is taken as the average of the magnitudes of the $\Delta_1$'s in the 
two phases, with zero phase factor.  With these assumptions, the dispersion is a
unique function of $m$ and $n$.  Figures \ref{fig:7a} ($m=n=2$) and \ref{fig:7b}
($m=2$, $n=6$) provide representative illustrations of the complicated 
dispersion to be expected. There are $2(m+n)$ subbands, with very small 
dispersion along the $x$ direction, since $t\simeq 0$ in the flux phase.
Note that, because of this small dispersion along $X$, the dispersion along
$\Gamma\rightarrow Y$ is nearly the same as that along $X\rightarrow S$.
\par
Once again, the results do not greatly resemble the photoemission data.  
Presumably, this is because the domains are fluctuating dynamically\cite{SEK}.  
If photoemission from a static domain pattern could be observed, then the 
minigaps predicted here should be observed as long as the domains are nearly 
commensurate and 
disorder effects are small.  In particular, the dispersion of Fig. \ref{fig:7a} 
has a hole doping $x_c/2=0.144$, and should be similar to the $x=1/8$ phase of 
LSCO and La$_{2-x}$Ba$_x$CuO$_4$.  [Note that the experiments of Tranquada, et
al.\cite{Tran} determined the overall periodicity of the stripes, but only
{\it assumed}, on analogy with the nickelates, that the charge stripes were one 
cell wide, corresponding to a hole doping $x=0.5$.  An equally good case can be
made\cite{Surv} for the assumption that the charge stripes are two cells wide,
with $x\simeq 0.25$.  This is consistent with the results of White and
Scalapino\cite{WhiSc}.]

\subsection{Dynamic Stripes}

\begin{figure}
\leavevmode
   \epsfxsize=0.33\textwidth\epsfbox{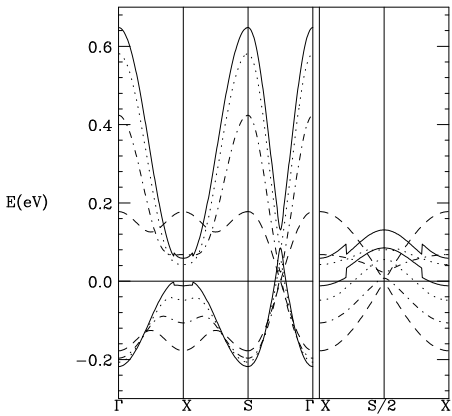}
\vskip0.5cm 
\caption{Energy dispersion for the fluctuating
stripe phase model, for $\nu_c$ = 0 (dashed line), 0.5 (dotdashed line),
0.75 (dotted line), and 1 (solid line).}
\label{fig:8}
\end{figure}
\par
Notice that once nanoscopic stripes are formed, the averaging inherent in Eq.
\ref{eq:12} is lost: no domain is large enough to have the dispersion
characteristic of $G_1$.  Now the band parameters have become local functions of
space and, in dynamic stripes, of time as well.  This can best be thought of as 
a generalization of the zero sound modes of a Landau Fermi liquid -- 
shape oscillations of the Fermi surface due to electron-electron interaction.
I propose that in this state, when the stripe motion is rapid enough that only
average properties are apparent, the appropriate procedure (replacing Eq.
\ref{eq:12}) is to {\it average the band parameters themselves}.
\par
This is done in Fig. \ref{fig:8}, for several intermediate dopings.  The
resulting dispersions are in good agreement with the experimental data, Fig.
\ref{fig:6}. The photoemission studies find that at optimal doping, the Fermi 
level is close to an {\it extended} VHS, which evolves into a bifurcated VHS in 
underdoped Bi-2212 (for a bifurcated VHS, the 
dispersion has two minima offset from the $X$ point along the $\Gamma -X$ line 
of the Brillouin zone).  These results are well reproduced in the present 
calculations.
\begin{figure}
\leavevmode
   \epsfxsize=0.33\textwidth\epsfbox{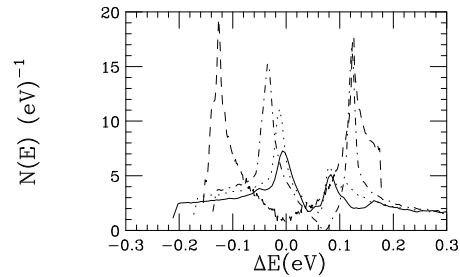}
\vskip0.5cm 
\caption{Density of states vs. energy $\Delta E=E-E_F$ for the fluctuating
stripe phase model, for $\nu_c$ = 0 (dashed line), 0.5 (dotdashed line),
0.75 (dotted line), and 1 (solid line).}
\label{fig:9}
\end{figure}

Figure \ref{fig:9} shows the calculated densities of states, 
illustrating the splitting of the VHS degeneracy.  Note that in the unmodulated 
phase at $x=x_c=0.288$, the VHS is split, but there is no true gap (with $N(E)=
0$).  This is because the $\delta_0$ gap has a dispersion, so parts of the Fermi
surface remain ungapped, at least in the absence of superconductivity.  For 
lower doping, the effective Fermi surface becomes closer to a square, and there 
is a gap over the full Fermi surface.

Given the large splitting of the VHS degeneracy near the $X$-point, one would
expect a large interband absorption associated with inter-VHS scattering.  If
it is recalled that the experimental dispersion is about a factor of two
larger than calculated (see Fig.~\ref{fig:6}), then this absorption could
readily be identified with the well-known mid-infrared absorption in the
cuprates\cite{Timu}.  This feature displays a considerable shift as a function
of hole doping, being centered at about 0.5eV at very low dopings, while 
at optimal doping, the peak has moved to $\sim$0.1eV.  The present
interpretation of the splitting would be consistent (at least near optimal
doping) with the model of ``electronic polarons'' analogous to Zhang-Rice 
singlets\cite{ElPo}.  In the low doping regime, there may be additional  
absorption peaks associated with an isolated charge stripe, as suggested by
the large number of very flat bands along $\Gamma\rightarrow X$ in Figs.
\ref{fig:7a},\ref{fig:7b}.

\begin{figure}
\leavevmode
   \epsfxsize=0.33\textwidth\epsfbox{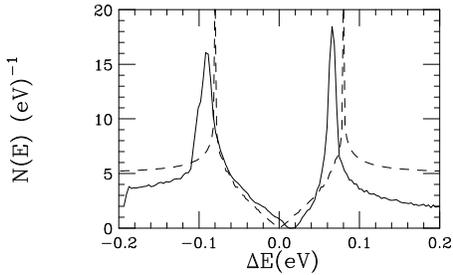}
\vskip0.5cm 
\caption{Density of states vs. energy $\Delta E=E-E_F$ for the fluctuating
stripe phase model, for $\nu_c$ = 0.5 (solid line), compared to the calculated
dos for a d-wave gap (dashed line).}
\label{fig:10}
\end{figure}

The opening of the pseudogap with underdoping found in Fig. \ref{fig:9} is in 
good qualitative agreement with the heat capacity measurements of Loram, et 
al.\cite{Gp3,Gp6}.  They have fit the dos lineshape to a d-wave gap, and 
Figure \ref{fig:10} shows that the dos calculated for $\nu_c=0.5$ does indeed 
bear a strong resemblance to a d-wave gap (dashed line).
\begin{figure}
\leavevmode
   \epsfxsize=0.33\textwidth\epsfbox{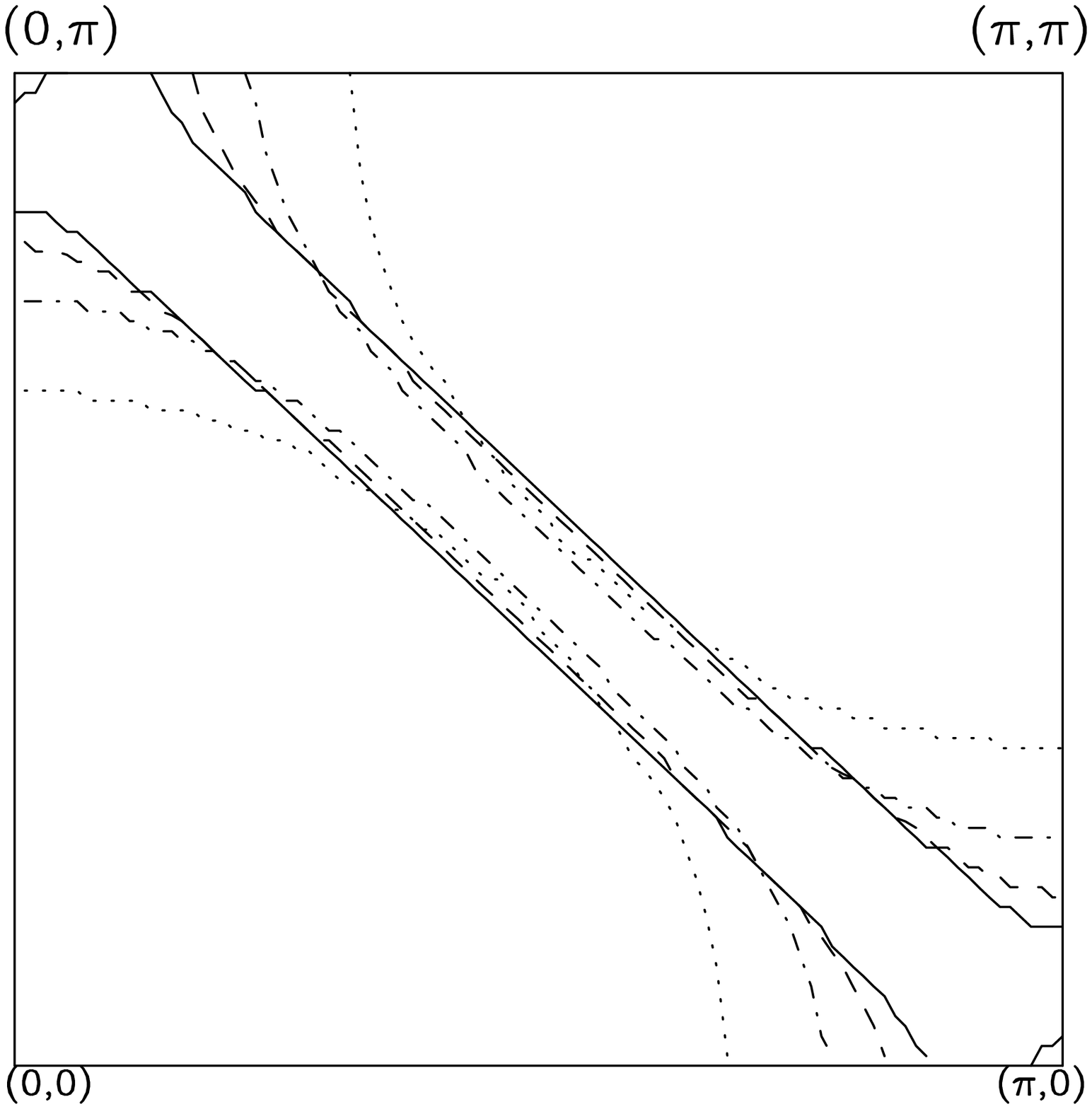}
\vskip0.5cm 
\caption{Fermi surfaces for the fluctuating stripe phase model, for $\nu_c$ = 
1.0 (dotted line), 0.625 (dotdashed line), 0.5625 (dashed line), and 0.5486 
(solid line).}
\label{fig:11}
\end{figure}
\begin{figure}
\leavevmode
   \epsfxsize=0.33\textwidth\epsfbox{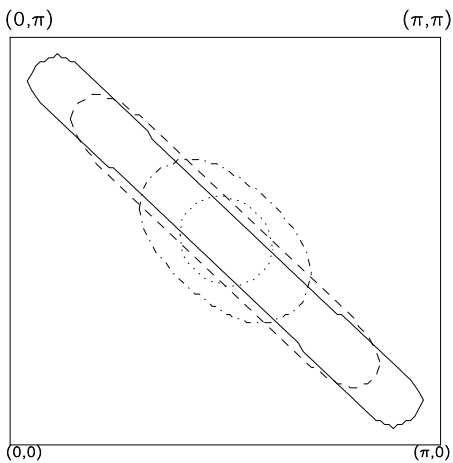}
\vskip0.5cm 
\caption{Fermi surfaces for the fluctuating stripe phase model, for $\nu_c$ = 
0.125 (dotted line), 0.25 (dotdashed line), 0.5 (dashed line), and 0.5469 
(solid line).}
\label{fig:12}
\end{figure}
\par
A closer comparison of the data, Fig. \ref{fig:6}, and the theory, Fig. 
\ref{fig:8}, reveals excellent agreement near $\Gamma$ (e.g., along $\Gamma
\rightarrow X$), but an absence of the experimental points near $S =(\pi ,\pi )$
(e.g., along $X\rightarrow S$).  This is the `ghost' part of the dispersion,
and it is found to be weak even in the limiting case of SCOC (squares in Fig. 
\ref{fig:6}).  The evolution of the Fermi surface with doping shows a clear
crossover from a large Fermi surface, Fig. \ref{fig:11}, to a small Fermi
surface, Fig. \ref{fig:12}.  Once again, the shapes of the Fermi surfaces are
in good agreement with photoemission, except in the ghost part of the 
dispersion.  
\par
Note that while the theoretical data in Figs. \ref{fig:0a}-\ref{fig:6}
are based on the self-consistent slave boson calculations of the flux and 
paramagnetic phases or on the result of a more exact calculation\cite{HM,Man}
(the point marked $\times$ in Fig. \ref{fig:0a}), 
the results of this section, Figs. \ref{fig:8}-\ref{fig:12}, require one 
additional assumption.  This is the assumption that, in the presence of a 
dynamic stripe phase, the proper average over these stripes is given by the 
weighted average of the band parameters of the two coexisting phases.  The
present results are consistent with the findings of Salkola, et al.\cite{SEK}.
They calculated the photoemission spectra of a phenomenological stripe phase
Hamiltonian, and found that (1) for a regular stripe array, the dispersion is
dominated by minigaps, and bears little resemblance to experiment, and (2) for
a quenched random distribution of stripes (taken as representative of 
dynamically fluctuating stripes) there is a single average photoemission
spectrum, which resembles the experimental observations.

\section{Discussion}

The above calculations have provided a plausible explanation for the opening of
a double pseudogap in underdoped Bi-2212, while providing a sounder underpinning
for the earlier one-band calculations of Ref. \cite{RM8a}.  The present 
results are in excellent agreement with these earlier calculations, which can
therefore be used to supplement the present T=0 calculations. Here, the detailed
correspondence between experiment and theory will be briefly summarized, 
including calculations of superconducting properties and temperature dependences
from the one-band model.  
\par
First and foremost, it should be noted that
Fig. \ref{fig:9} is a very clear demonstration that the physics of the
underdoped cuprates is dominated by VHS nesting, with the splitting of the dos
peak increasing smoothly with increased underdoping.  Secondly, the model is not
merely consistent with the underlying presence of striped phases (or similar
manifestations of nanoscale phase separation), it {\it requires} them to
reproduce the smooth doping dependence of the pseudogap magnitude.  Thirdly,
it correctly reproduces the characteristic double gap structure which has been 
such a puzzling feature of the photoemission experiments: the large splitting
of the VHS degeneracy, associated with $\delta_0$, and the smaller pulling back 
of dos from the Fermi level, associated with $\delta_1$.  Since both are aspects
of a single transition, both disappear at the same pseudogap transition
temperature, as found experimentally.  
\begin{figure}
\leavevmode
   \epsfxsize=0.33\textwidth\epsfbox{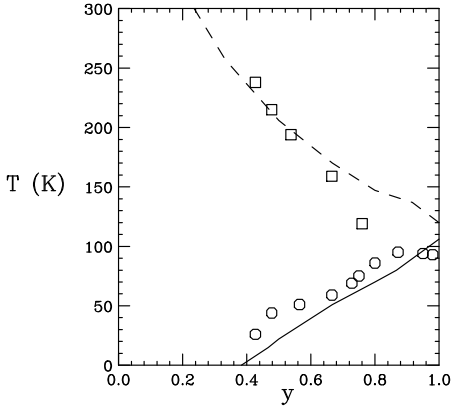}
\vskip0.5cm 
\caption{Phase diagram of the pseudogap (dashed line) and superconducting (solid
line) phases, based on the one-band model, and compared to the experimental data
of Loram, et al.\protect\cite{Lor} for YBa$_2$Cu$_3$O$_{6+y}$.}
\label{fig:13}
\end{figure}
\par
Figure \ref{fig:13}\cite{RMPRL} shows that the model reproduces the 
experimentally observed\cite{Lor} pseudogap phase diagram, including a doping 
dependent superconducting $T_c$. The competition between pseudogap and 
superconductivity displayed
in Fig. \ref{fig:13} can be understood in terms of the evolution of the shape
of the Fermi surface with doping.  The VHS is pinned by correlation and phase
separation effects near the Fermi surface; this pinning means that the shape of
the Fermi surface must evolve with doping, being square at half filling, and
curved in such a way as to accomodate more holes as doping increases.  Since the
pseudogap is associated with nesting, it dominates near half filling, when the
nesting is perfect.  Away from half filling, the nesting is worse, so the
pseudogap and pseudogap transition temperature both rapidly decrease with
doping.  Superconductivity requires a large dos, but is insensitive to nesting;
moreover, it can arise on those sections of Fermi surface which survive the 
imperfect nesting.  For both of these reasons, the superconducting transition
temperature grows with increasing doping, until it is comparable to the
density wave transition.  

\begin{figure}
\leavevmode
   \epsfxsize=0.33\textwidth\epsfbox{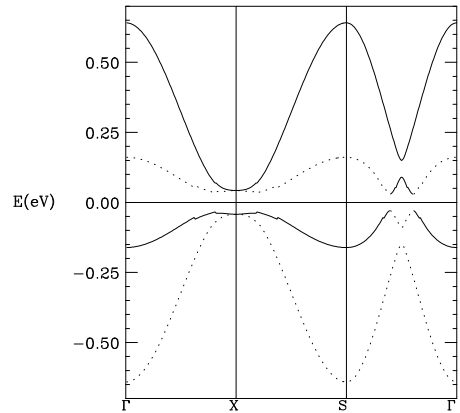}
\vskip0.5cm 
\caption{Energy dispersion in the one-band model, in the presence of
superconductivity.  Dotted lines = hole-like part of dispersion (electron-like
part reflected around Fermi level).  Only the electron-like part (solid lines)
will be visible in, e.g., a photoemission experiment.  To stress the role of the
superconducting gap, the smaller CDW gap, $\delta_1$, has been set equal to 
zero.}
\label{fig:14}
\end{figure}

Within the model, the pseudogap arises from a structural or magnetic
instability -- actually, there is a crossover between the two effects with 
increased underdoping.  Neither has any direct relation with superconductivity, 
except that they compete with it for the large dos associated with the VHS.
Superconductivity first arises in the model from the leftover dos associated
with parts of the Fermi surface away from the VHS.  This is illustrated in
the one-band model calculation of Fig. \ref{fig:14}, which shows how the
electronic dispersion, in the presence of a density wave gap, is modified by
the appearence of a superconducting gap.  [For simplicity, an s-wave gap is
assumed, rather than the more realistic d-wave gap.]  In this figure, the 
hole-like parts of the Fermi surface (dotted lines) are ghosts, with the 
intensity suppressed by coherence
factors, and will not be observed by photoemission.  The solid lines in this
figure should be compared to the dotted lines in Fig. \ref{fig:8}.  The main
difference is the superconducting gap away from the VHS -- mainly near the 
hole pockets at $S/2$, although there is a weak superconducting contribution to 
the gap at the VHS.  Thus, unlike Fig. \ref{fig:9}, in the presence of a 
superconducting gap, the dos will vanish at the Fermi level.

\par
Hence, in the present model, the smaller photoemission gap near the Fermi
level is a composite object, due in part to superconductivity but also in part
to the density wave gap.  This explains why a gap persists near $(\pi ,0)$ in 
the pseudogap regime above the superconducting $T_c$.  However, at $T_c$ there 
should be subtle changes in the form of the gap -- in particular above $T_c$, 
the density wave gaps should vanish in a finite portion of the Brillouin zone 
near $S/2$.  The present model would predict a scaling of the superconducting 
part of the gap with $T_c$, and hence with $\nu_c$\cite{RM8a}.  Evidence for 
such a gap feature is found in neutron scattering measurements of the
magnetic susceptibility near $S=(\pi ,\pi )$, which see a spin gap followed by a
resonance peak, both of which features scale with $T_c$\cite{spgp}.  Such a
spin gap follows from BCS theory.  To be consistent with Fig. \ref{fig:14}, the
susceptibility must be associated with scattering between sections of Fermi
surface near $S/2$ which are not gapped by the density wave order.
\par
Within the present model, the issue of {\it overdoping} can be briefly
addressed.  From the free energy curves of Fig. \ref{fig:1}, it appears that
the system will continue to evolve in a uniform paramagnetic phase for doping
beyond the phase separated regime.  However, the free energy has a local minimum
near optimal doping, which is likely to be enhanced by the difficulty of adding
additional holes to the CuO$_2$ planes.  Hence, there may well be another phase
with lower free energy, probably associated with doping holes off of the
planes (perhaps onto the apical O's and Cu-d$_{z^2}$'s) in this doping range.
The crossover from the optimally doped phase to this overdoped phase will 
probably again involve a phase separation.  Experimental evidence for a second 
phase separation in the overdoped regime is summarized in Ref. \cite{Surv}, 
Section 11.6.  This does not preclude the possibility that there is a small but
finite range of doping near the optimal in which a single phase solution is
stable.  
Whereas in YBCO the pseudogap and superconducting transition temperatures
coincide at optimal doping, in LSCO the pseudogap temperature is considerably 
higher than $T_c$, even at optimal doping\cite{Surv}.
Interestingly, heat capacity measurements in LSCO\cite{Gp3} find that
the gap closes to a single VHS peak at the Fermi level (see Fig. 21 of Ref.
\cite{Surv}) in the overdoped range, $x$=0.27.  For larger overdopings, this
peak remains at the Fermi level, but decreases in intensity (as might be
expected in the presence of a phase separation).  This suggests an
even stronger pinning of the Fermi level to the VHS than expected theoretically.
Note from Fig. \ref{fig:1} that in the overdoped regime, the CDW undergoes a 
quantum phase transition, $T_{CDW}\rightarrow 0$.  The possible role of such
a QCP on superconductivity has been discussed recently\cite{PCCG}.  However, 
this could be obscured by a second phase separation in the overdoped regime.
\begin{figure}
\leavevmode
   \epsfxsize=0.33\textwidth\epsfbox{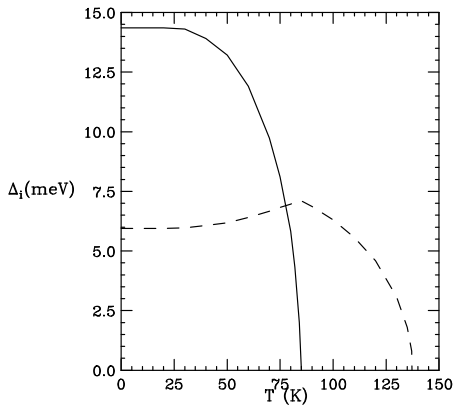}
\vskip0.5cm 
\caption{Temperature dependences of the pseudogap (dashed line) and the
superconducting gap (solid line) in the one-band model.}
\label{fig:15}
\end{figure}
\par
There have been a number of alternative interpretations of the pseudogap.  The
strictly magnetic models\cite{SFNL} have difficulty explaining why a gap is
also seen in the charge spectrum, including photoemission and heat capacity.
Other models suggest that it is associated with local pair formation, as a 
precursor effect to superconductivity\cite{LPs}.  However, in overdoped 
materials, the pseudogap transition lies at a lower temperature than the
superconducting transition\cite{ovps}.  Moreover, these models do not explain 
the frequent association of the pseudogap with 
structural anomalies, Ref. \cite{Surv}, Sections 9.1,2.  Perhaps the clearest 
example is in YBa$_2$Cu$_4$O$_8$. Even when stoichiometric, this material is 
underdoped, behaving in many ways like YBa$_2$Cu$_3$O$_{6.6}$, with a pseudogap 
onset near 150K.  When some of the Y is replaced by Ca, a transition to a 
long-range structually ordered phase is found at nearly the same transition 
temperature\cite{TF3}.  Moreover, the theory predicts that, in the
doping range where both phases coexist, the onset of superconductivity leads to
a softening of the pseudogap, Fig. \ref{fig:15}.  This can explain a number of
observations of lattice anomalies at T$_c$, Ref. \cite{Surv}, Section 9.3.
\par
While the present model provides an impressive picture for pseudogap formation
in the presence of dynamic stripes, it must be
recalled that a number of intermediate steps need to be filled in.  These
include: (1) self consistent calculation of the
band parameters in a static striped phase; and (2) incorporation of dynamical
fluctuations into the calculation.  In addition, (3) a more detailed analysis of
just which phonon modes are coupled is necessary, to see how both the smaller
DW gap and the superconducting gap can be d-wave.  Despite these limitations, it
is clear that the present calculations have the ability to explain both the 
stripes and the pseudogap within a common theoretical framework.

\end{document}